\documentclass[12pt]{article}
\usepackage[reqno]{amsmath}
\usepackage{amsthm}
\usepackage[dvips]{epsfig,color,graphicx}
\usepackage{amssymb} 
\def\be{ \begin{equation}}          \def\ee{ \end{equation}}
\def\ba{ \begin{eqnarray}}          \def\ea{ \end{eqnarray}}

\def\nn{\nonumber}                  

 \def\Z{\mathbb{Z}} \def\R{\mathbb{R}}

\def\cedille#1{\setbox0=\hbox{#1}\ifdim\ht0=1ex \accent'30 #1%
 \else{\ooalign{\hidewidth\char'30\hidewidth\crcr\unbox0}}\fi}
\def\gaw{Gaw\cedille edzki}

\def\Ad{\mbox{\rm Ad}}

\def\sg{{\cal G}}

\def\asg{{\widehat{\sg}}}

\def\a{\alpha }

\def\Mat{{\mbox{\rm Mat}}}

\def\cH{{\cal H}}

\def\tr{{\rm tr}}
\def\cS{{\cal S}}
\def\tF{{\rm F}}
\def\tCS{{\rm CS}}
\def\tf{{f}}
\def\tL{{\rm L}}

\def\tA{{\rm A}} 
\def\astk{\, , \, }

\title{\bf Brane Dynamics in CFT Backgrounds \footnote{Talk given by 
V.\ Schomerus at Strings 2001, Mumbai, India.}}
\author{{\sc Stefan Fredenhagen {\rm and} Volker Schomerus}\\[9mm]  
     Max-Planck-Institut f\"ur Gravitationsphysik,\\ 
       Albert-Einstein-Institut, Am M\"uhlenberg 1,\\ D--14424 Potsdam,
        Germany}
\date{April 4, 2001}
\begin{document}
\maketitle      

\vskip.5cm
\begin{abstract}
In this note we discuss bound states of un- or meta-stable brane 
configurations in various non-trivial (curved) backgrounds. We 
begin by reviewing some known results concerning brane dynamics 
on group manifolds. These are then employed to study condensation in 
cosets of the WZW model. While the basic ideas are more general, our 
presentation focuses on parafermion theories and, closely related, 
$N=2$ superconformal minimal models. We determine the (non-commutative) 
low energy effective actions for all maximally symmetric branes in a 
decoupling limit of the two theories. These actions are used to show 
that the lightest branes can be regarded as elementary constituents 
for all other maximally symmetric branes.
\noindent  
\end{abstract}
\vspace*{-15.9cm}
{\tt {hep-th/yymmxxx  \hfill AEI 2001-26}} \vspace*{15.9cm}

\section{Introduction} 

Many aspects of brane physics in string compactifications have 
been studied in a large volume regime using geometrical methods. 
When the volume becomes small, however, there can be strong 
string corrections causing the classical description to fail.  
Exact results are then to be based on a `microscopic' 
approach to D-branes which makes use of boundary conformal 
field theory.  
\smallskip

Boundary conformal field theory offers powerful techniques for the 
construction of branes, i.e.\ of their couplings to closed string 
modes, their open string spectra and scattering amplitudes. For the 
purpose of this note we shall assume that we are given the boundary 
conformal field theory of some brane configuration. The latter 
may be un- or metastable in which case our boundary conformal 
field theory contains relevant or marginally relevant boundary 
fields. Such fields can generate renormalization group (RG) flows 
into new boundary conformal field theories that are associated with 
the stable decay product (or bound state) of the original brane
configuration. These processes are the main subject of this note. 
\medskip

Typically, renormalization group flows are quite difficult to 
analyse. Many models, however, possess some limit in which the 
end-point of the RG flow lies arbitrarily close to the starting 
point (in the space of (renormalized) boundary couplings). 
In such cases, the flow between the two fixed points can be 
studied perturbatively. For boundary conformal field theory, 
the first investigations of this type were performed in 
\cite{ReRoSc}. 
\smallskip

The limiting regime in which perturbative studies yield 
reliable results on RG flows is similar to the decoupling 
limit of string theory and, in general, it admits for an 
effective description through some non-commutative world-%
volume field theory as in \cite{SeiWit}. An illuminating
example is provided by the relation between brane 
dynamics on group manifolds and gauge theory on a fuzzy 
space \cite{AlReSc2}. 
\smallskip

Strings and branes on group manifolds are analysed with the 
help of the WZW model. The latter is probably the most 
important ingredient in CFT model building. Hence, one may 
hope that a good understanding of brane dynamics on group 
manifolds has direct implications on other backgrounds. It 
is one purpose of this note to demonstrate this explicitly 
at the example of parafermion theories and $N=2$ minimal 
models. These two theories are both obtained as cosets of 
the SU(2) WZW-model.
\medskip

Below we shall begin with a reminder on branes and brane dynamics
in the group manifold SU(2)$\cong S^3$. Section 2 deals with 
semi-classical aspects of brane geometry in SU(2) \cite{AlSc1}
on which we then build our review of the decoupling limit 
in Section 3. After presenting the non-commutative gauge theory 
that describes branes on SU(2), we recall from \cite{AlReSc2} that 
several point-like branes on SU(2) are metastable against decay into 
a single spherical brane. In Section 4 we shall briefly discuss how 
such processes may be analysed beyond the decoupling limit 
\cite{AlSc2,FreSch}. Finally, in Section 5, we reduce the action of 
branes on SU(2) to study bound state formation in parafermion 
theories and $N=2$ minimal models. This last section contains a 
number of new results.

\section{Semi-classical geometry of branes on $S^3$} 
\def\rmA{{\rm A}}

Strings moving on a 3-sphere $S^3 \cong$ SU(2) of radius 
$R \sim \sqrt{k}$ are described by the SU(2) WZW model 
at level $k$. The string equations of motion imply that 
the 3-sphere comes equipped with a constant NSNS 3-form 
field strength $ H \sim \Omega$ where $\Omega$ 
denotes the volume form of the unit sphere. In the present 
context $H$ is also known as the WZW 3-form. 
\medskip

The world-sheet swept out by an open string in $S^3$ is 
parametrised by a map $g: \Sigma \rightarrow {\rm SU(2)}$ 
from the upper half-plane $\Sigma$ into the group manifold 
SU(2)$\, \cong S^3$.  We shall be interested in maximally 
symmetric D-branes on SU(2) which are characterised by 
imposing the condition  
\begin{equation}  - k \, (\partial g) g^{-1} \ = :\ J(z) \ 
 \stackrel{!}{=} |_{\ _{\!\!\! z =\bar z}}  
 \bar J(\bar z) \ := \ k \, g^{-1} \bar\partial g\  
\label{GC}  \end{equation}
on chiral currents all along the boundary $z = \bar z$. With 
this choice of boundary conditions the theory can be solved 
exactly using purely algebraic methods of boundary conformal 
field theory. Even though much of the construction was known 
for more than 10 years \cite{Car}, the brane geometry encoded 
in rel.\ (\ref{GC}) was only deciphered more recently \cite{AlSc1}.    
\smallskip

Formally, the relation (\ref{GC}) looks similar to Dirichlet
boundary conditions for branes in flat space (note that there
appears an extra minus sign in the definition of $J(z)$). But 
it turns out that this is not the correct answer. To describe 
the findings of \cite{AlSc1} we decompose the  tangent space 
$T_g = T_g{\rm SU}(2)$ at each point $g \in \,$SU(2) into a 
part $T^{||}_g$ tangential to the conjugacy class through $g$ 
and its orthogonal complement $T^\perp_g$ (with respect to the 
Killing form). It is then easy to see that eq.\ (\ref{GC})
implies  
$$ (g^{-1} \partial_x g)^\perp \ = \ 0 \ \ .$$
In other words, the string ends can only move within the 
conjugacy classes on SU(2). Except for two degenerate cases, 
namely the points $e$ and  $-e$ on the group manifold, these 
conjugacy classes are 2-spheres. These branes carry a 
non-vanishing B-field that can also be read off from eq.\ 
(\ref{GC}). It has the form    
\begin{equation} B \ \sim  \tr \left( g^{-1} dg \, 
    \frac{\Ad(g)+1}{\Ad(g)-1}\, g^{-1} dg \right)  \ \ ,  
\label{Bfield} \end{equation} 
where $\Ad(g)$ denotes the adjoint action of $G$ on its Lie 
algebra. The last two formulas hold for arbitrary groups 
and one can show in the general case that $B$ provides a
2-form potential for the WZW 3-form $H$. It was argued in 
\cite{BaDoSc}, \cite{Paw} that the spherical branes are 
stabilised by the NSNS background field $H$.

\section{Decoupling limit and fuzzy gauge theories}

As in the case of branes in flat space (see \cite{Vol} and references 
therein), the presence of the B-field implies that the brane comes 
equipped with some bi-vector $\Theta$. The latter defines a Poisson 
structure in the limit $k \to \infty$ where the 3-sphere grows and 
approaches flat 3-space $\R^3$. \footnote{For finite $k$, 
$\Theta$ does not obey the Jacobi identity.} It was found 
in \cite{AlReSc1} that this Poisson structure on $\R^3$ 
is linear, i.e. that     
\begin{equation} 
\label{LPB1}
  \Theta_{ab} \ = \ f_{ab}^{\ \ c}\, y_c \ \ ,   
\end{equation}     
in terms of the three coordinate functions $y_c$ on the 
3-dimensional flat space. 
\medskip

Recall that the Moyal-Weyl products that show up for brane geometry 
in flat space with constant B-field are obtained from the constant 
Poisson structure $\Theta_{\mu\nu}$ on $\R^d$ through quantisation. 
Consequently, we expect that the quantisation of 2-spheres in $\R^3$ 
with Poisson structure  (\ref{LPB1}) becomes relevant for the geometry 
of branes on ${\rm SU(2)}$ in the limit where $k \rightarrow \infty$. 
This quantisation problem has been addressed from various angles and
the solution is well known. 
\smallskip

It implies that only a discrete set of 2-spheres in $\R^3$ can be 
quantised with their radii being related to the values of the quadratic 
Casimir for the irreducible representations of ${\rm su(2)}$. Hence, the 
quantisable spheres are labelled by one discrete parameter $\a = 0, 
1/2, 1, \dots $. For each quantisable 2-sphere $S^2_\a \subset \R^3$ 
one obtains a state space $V^\a$ of finite dimension $\dim V^\a = 2\a + 1$ 
equipped with  an action of the quantised coordinate functions $\hat y_c$ 
on $V^\a$. The latter represent the generators $t_c$ of su(2) in the 
representation $D^\a$ on $V^\a$, i.e.\ $\hat y_c = D^\a(t_c)$,  and they 
generate the matrix algebra $\Mat (2\a +1)$ which is also known as a 
fuzzy 2-sphere \cite{HopMad}.
\medskip 

Open string amplitudes for branes on group manifold can be computed
from the exact solution \cite{Car,Run,AlReSc1} of the boundary WZW 
model. Note that the knowledge
of the propagator is not sufficient because we are dealing with an 
interacting field theory in which Wick's theorem does not hold. 
Otherwise, the computation of the effective action follows the 
same steps as in \cite{SeiWit} with two important changes: First, 
as we argued above, the Moyal-Weyl products are to be replaced 
by matrix products. The size of the matrices corresponds to the 
radius of the brane that we want to describe. Moreover, there 
appear some extra terms in the computation which are proportional 
to the structure constants $f_{ab}^{\ \ c}$. They give rise to 
a Chern-Simons like term in the effective action.
\smallskip

For $Q$ branes of type $\a$ on top of each other, the results 
of the complete computation \cite{AlReSc2} can be summarised in the 
following formula, 
\begin{equation}  \label{effact}
  \cS_{(Q , \a)}\ =\  \cS_{{\rm YM}} + \cS_{{\rm CS}}\ = \
      \frac{1}{4}\ \tr \left( \tF_{a b} \ \tF^{a b} \right) 
             - \frac{i}{2}\  \tr \left( \tf^{a b c}\; \tCS_{a b c} \right)
\end{equation}
where we defined the `curvature form' $\tF_{ab}$ and some 
non-commutative analogue $\tCS_{abc}$ of the Chern-Simons form
by the expressions     
\begin{eqnarray} \label{fieldstr}
  \tF_{a b}(\tA) & = &    
   i\, \tL_a \tA_b - i\, \tL_b \tA_a + i \,[ \tA_a \astk \tA_b] 
   +  \tf_{a b c } \tA^c \\[2mm]
   \label{CSform}
  \tCS_{a b c}(\tA) & = & \tL_a \tA_b \, \tA_c 
                   + \frac{1}{3}\; \tA_a \, [ \tA_b \astk \tA_c]
                   - \frac{i}{2}\; \tf_{a b d}\; \tA^d \, \tA_c \ \ .
\end{eqnarray}
The three fields $\tA_a$ on the fuzzy 2-sphere $S^2_\a$ may take 
values in $\Mat(Q)$ for some Chan-Paton number $Q \geq 1$ meaning 
that the $\tA_a$ are elements of $\Mat(Q) \otimes \Mat(2\a+1)$. We 
also introduced the symbol $\tL_a$ to denote the `infinitesimal 
rotation' $\tL_a \tA = [ {\bf 1}_Q \otimes \hat y_a, \tA]$ acting on 
arbitrary elements $\tA \in \Mat(Q) \otimes \Mat(2\a+1)$. Gauge 
invariance of (\ref{effact}) under the gauge transformations 
$$ \tA_a  \ \rightarrow \ \tL_a \Lambda \ +\  i\, [\, \tA_a\, ,\, 
    \Lambda\, ]  \ \ \ \mbox{ for } \ \ \ \Lambda \in 
    \Mat(Q) \otimes \Mat(2\a+1)  
$$
follows by straightforward computation. Note that the 'mass term' 
in the Chern-Simons form (\ref{CSform}) guarantees the gauge 
invariance of $\cS_{{\rm CS}}$. On the other hand, the effective 
action (\ref{effact}) is the unique combination of $\cS_{\rm YM}$ 
and $\cS_{\rm CS}$ in which mass terms cancel. 
\medskip

Stationary points of the action (\ref{effact}) describe condensates 
on a stack of $Q$ branes of type $\a$. A simple analysis reveals 
that there is an interesting set of such classical solutions that 
has no analogue for flat branes in flat backgrounds. In fact, any 
$Q(2\a + 1)$-dimensional representation of the Lie algebra su(2) 
lies in a local minimum of the action (\ref{effact}). Their  
interpretation was found in \cite{AlReSc2}. For simplicity we 
restrict our discussion to a stack of $Q$ point-like branes 
($\a = 0$) at the origin of SU(2). In this case, 
$\tA_a \in \Mat(Q) \otimes \Mat(1) \cong \Mat(Q)$ so that we need a $Q$-%
dimensional representation of su(2) to solve the equations of motion. 
Let us choose the $Q$-dimensional irreducible representation $\sigma$. 
Our claim then is that this drives the initial stack of $Q$ point-like 
branes at the origin into a final configuration containing only a 
single brane wrapping the sphere of type $\a = (Q-1)/2$, i.e.\ 
$$ (Q,\,\a = 0 )  \ \stackrel{\sigma}{\longrightarrow} \ 
   (1,\,\a = (Q-1)/2) \ \ . $$
Support for this statement comes from the comparison of tensions and the
fluctuation spectrum (see \cite{AlReSc2}).
Similar effects have been described for branes in RR-background 
fields \cite{Mye}. The advantage of our scenario with NSNS-background
fields is that it can be treated in perturbative string theory so 
that string effects may be taken into account. 

\section{Dynamics in stringy regime and K-theory}
 
Now we would like to understand the dynamics of branes in the stringy 
regime. Proceeding  along the lines of the previous subsection would 
force us to include higher order corrections to the effective 
action. Unfortunately, such a complete control of the brane dynamics 
in the stringy regime is out of reach.  
\smallskip

But we could be somewhat less ambitious and ask whether at least some 
of the solutions we found in the large volume limit possess a deformation 
into the small volume theory and if so, which fixed points they 
correspond to. It turns out that this is possible for all the processes 
that are obtained from constant gauge fields on the brane. In fact, 
constant condensates on branes on group manifolds are very closely 
related to the low temperature fixed point of Kondo models. Consequently, 
the analysis of such condensation processes in the stringy regime can 
be based upon very thorough renormalization group studies that go even
back to the work of Wilson. 
\smallskip

As an aside, let us briefly discuss how the Kondo problem and
the study of constant gauge field condensates translate into 
each other \cite{AffLud}. The Kondo model is designed to understand 
the effect of magnetic impurities on the low temperature conductance 
properties of a conductor. The latter can have electrons in 
a number $k$ of conduction bands. We can build several currents 
from the basic fermionic fields. Among them is the spin current
$\vec{J}(t,y)$ which gives rise to a $\asg_k$ current algebra. The 
coordinate $y$ measures the radial distance from a spin $s$  
impurity at $y=0$ to which the spin current couples. This coupling 
involves a $2 s + 1$-dimensional irreducible representation 
$\vec{\Lambda} = (\Lambda_a, a=1,2,3)$ of su(2) and it is of 
the form 
\begin{equation} \label{Kondo} 
    S_{\rm pert} \ \sim \ \int_{-\infty}^\infty {\rm d}t \, \Lambda_a J^a(t,0) 
    \ \ . 
\end{equation}
This term is identical to the coupling of open string ends to a 
background gauge field $\rmA_a = \Lambda_a \in \Mat(2s+1)$. Hence, 
$\Lambda_a$ may be interpreted as a constant gauge field on a 
Chan-Paton bundle of rank $2s+1$.      
\medskip

Now let us consider a supersymmetric theory on a finite 3-sphere 
with $K = k+2$ units of NSNS flux passing through. In this case
one can have (anti-)branes wrapping $k+1$ different integer 
conjugacy classes labelled by $\a = 0,1/2, \dots,k/2$ (see e.g. 
\cite{AlSc2}). As in the limit of infinite level $k$, we consider 
a stack of $Q$ point-like branes at the origin and a field $\rmA_a 
= \Lambda_a \in \Mat(Q)$ whose components give rise to an 
irreducible representation of su(2). Then renormalization group 
studies show that this stack will decay into a single object
wrapping the conjugacy class $\a = (Q-1)/2$ on $S^3$.   
\smallskip

\begin{figure}{ 
\begin{center}
\scalebox{0.8}[0.2]{\input{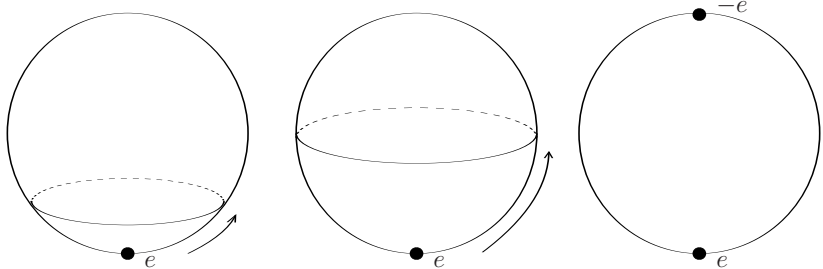}}
\end{center}
\caption{\label{su2branen} Brane dynamics on $S^3$: 
A stack of  point-like branes at $e$ can decay into a single spherical 
object. The distance of the latter increases with the number of branes
in the stack until one obtains a single point-like object at $-e$.}}
\end{figure}

If we stack more and more point-like branes at the origin the radius of 
the sphere that is wrapped by the resulting object will first grow, 
then decrease, and finally a stack of $k+1$ point-like branes at $e$ 
will decay to a single point-like object at $-e$ (see fig.~\ref{su2branen}).  
By taking orientations into account, one can see that the final
point-like object is the translate of an anti-brane at $e$. Hence, 
we conclude that the stack of $k+1$ point-like branes at $e$ has 
decayed into a single point-like anti-brane at $-e$. 
\smallskip

We would like to see whether the described brane dynamics obey some 
conservation laws, i.\,e.\ if we can assign charges to the branes that are
conserved in physical processes. In other words, we are looking for some 
discrete abelian group $C(X)$, where $X = (S^3,K)$ denotes the physical 
background, and a map from arbitrary brane configurations to $C(X)$ such 
that the map is invariant under renormalization group flows.
\smallskip

In our concrete example, we may assign charge~$1$ to the point-like
branes at $e$ and if we want the charge to be conserved, the decay 
product of $k+1$ such point-like branes must have charge $k+1$. On 
the other hand we identified the latter with a single anti-brane which 
has charge $-1$. Thus we have to identify $k+1$ and $-1$ which means 
that charge is only well-defined modulo $K=k+2$ \cite{AlSc2}, i.e.\ 
$$ C(SU(2),K) \ = \ \Z_K \ \ . $$
According to a proposal of Bouwknegt and Mathai \cite{BouMat}, the 
brane charges on a background $X$ with non-vanishing NSNS 3-form 
field $H$ take values in some twisted K-groups $K_H^*(X)$ which 
feel the presence of $H \in H^3(X,\Z)$. In most cases, these 
K-groups are difficult to compute. But for $S^3$ the answer is 
known to be $K^*_H(SU(2)) = \Z_K$. Here, $H$ is the $K^{th}$ 
multiple of the volume form $\Omega$ for the unit sphere. Hence, 
$C(SU(2),K) = K^*_H(SU(2))$ as predicted in \cite{BouMat}.        

\section{Brane dynamics in parafermion theories} 

The WZW model is probably the most important ingredient  
in conformal field theory model building through coset 
and orbifold constructions. Hence, one expects that the 
results for brane dynamics on group manifolds which we
reviewed above become relevant for many other string 
backgrounds. In this section we demonstrate how one may 
descent from the SU(2) WZW model to the coset SU(2)/U(1) 
describing parafermions. A more detailed discussion 
of such constructions will appear in a forthcoming 
paper.
\medskip

To begin with, let us briefly recall the coset construction 
SU(2)$_k/$U$_{2k}$ of parafermions. In the usual conventions,  
the numerator theory has sectors $\cH_{(l)}$ where $l = 
0,1, \dots,k$. We embed the free bosonic theory U$_{2k}$ such 
that its current gets identified with the component $J^3$ of 
the SU(2) current. The sectors of the denominator algebra 
U$_{2k}$ carry a label $m = -k+1, \dots , k$. According to 
the standard rules of the coset construction, we can label 
the sectors $\cH_{(l,m)}$ of the coset algebra by pairs 
$(l,m)$ such that the sum $l+m$ is even. 
\smallskip 

Our aim now is to select those states in the sectors $\cH_{(l,m)}$ 
of the parafermionic algebra that become massless as we send $k$ to 
infinity. We can find them within the subspace 
$$ \cH^{(1)}_{(l)} \ := \ \{ \, \psi \in \cH_{(l)}\, |\, \lim_{k
    \rightarrow \infty} \, h(\psi) \ = \ 1 \ \} $$   
of the sector $\cH_{(l)}$ for the SU(2) current algebra. Here, 
$h(\psi)$ denotes the conformal dimension of the state $\psi$.  
A careful analysis of the coset construction reveals that 
states $|\tA\rangle \in \cH_{(l)}^{(1)}$ obeying  
\begin{equation}
\label{restr}
(J^3_0 - m)\, |\tA\rangle \ =\ 0 \ \ \ \mbox{ and } 
\ \ \  J^3_1 \, |\tA\rangle \ = \ 0  \ \ 
\end{equation} 
give rise to marginal or marginally relevant perturbations in 
the $k \rightarrow \infty$ limit of parafermion theories. We will 
apply this result shortly after introducing the maximally symmetric 
branes in our coset theory.  
\smallskip

Cardy's construction provides us with a set of boundary conformal 
field theories which are in unique correspondence with sectors of 
the chiral algebra, i.e.\ they are labelled by $(L,M)$ being in the
same range as $(l,m)$ above. In the large $k$ limit, the tension 
$T_{(L,M)} = T_L$ of the associated branes is $L$ times the value 
of the tension $T_0$. This implies that the branes of type $(0,M)$ 
are the lightest in the theory. Moreover, one  can compute the 
open string spectrum for any theory $(L,M)$ from the fusion 
rules of the parafermion algebra. It contains the sectors 
$\cH_{(l,0)}$ where $l$ ranges from $0$ to $2L$, provided that 
$k \geq 2L$. Since all the sectors have $m=0$, the massless states 
in the large $k$ regime are found by imposing the conditions 
(\ref{restr}) with $m=0$. 

The effective field theory for these massless states of the 
brane $(L,M)$ can be obtained by a reduction from the effective 
theory (\ref{effact}) of a single brane $\a = L/2$ in the WZW 
model. To this end, we rewrite the constraints (\ref{restr}) 
with $m=0$ in terms of the world-volume fields $\tA_a$,  
\begin{equation}
\label{red}
\tL_3\tA_a \ =\ i\, f_{3a}^{\ \ b} \, \tA_b 
\ \ \ \mbox{ and } \ \ \ 
i\, f^{3a}_{\ \ b}\, \tL^b \tA_a + \frac{k}{2}\, \tA_3 \ =\ 0 
\ \  .
\end{equation}
For large $k$, the second equation becomes $\tA_3 = 0$ so that 
we can drop $A_3$ from the action (\ref{effact}). A short 
computation using the first equation in (\ref{red}) then gives 
the following effective action for the coset brane  
\begin{eqnarray} \label{redeffact}
\cS_{(L,M)}(\tA_1,\tA_2)  & = & \frac{1}{4} \, 
   \tr \left(\, \tF_{ij}\,  \tF^{ij}\, \right) \\[2mm]
 \mbox{where}\ \ \   \tF_{ij} & := & i\, \tL_i \tA_j - i \, \tL_j \tA_j 
          + i \, [\, \tA_i\, ,\, \tA_j\, ] \ \  \nn        
\end{eqnarray}     
up to terms of higher order in $1/k$. Here $i,j$ run through $1,2$ only 
and the matrices $\tA_1,\tA_2$ are elements of $\Mat(L+1)$.
This action still contains $2(L+1)^2$ parameters. Through the  
constraints (\ref{red}), we select a $2L$-dimensional subspace of 
physical parameters which correspond to marginally relevant 
perturbations of the brane in our coset theory. The action 
(\ref{redeffact}) together with  the constraints (\ref{red}) 
describes the dynamics of branes in the large $k$ regime of
parafermions.
\smallskip

It is rather easy to see that for $L>0$ the effective 
theory (\ref{redeffact}), 
(\ref{red}) has a stationary point at the following non-constant 
field
\begin{equation}\label{sol} 
\tA_i \ =\ - {\hat y}_i \ \ .
\end{equation}
Insertion of this solution into the effective action (\ref{redeffact})
gives a positive value, indicating that our brane $(L,M)$ is the decay 
product of some configuration with higher mass. We will shortly identify 
this configuration as a chain of branes 
\begin{equation} \label{chain} 
(0\, ,\, M-L)\, + \,  (0\, ,\, M-L+2)\, + \,  \dots \, + \, 
(0\, ,\, M+L)\ \ . 
\end{equation} 
Note that all the constituent branes $(0,M')$ have minimal tension. 
Evidence for our interpretation comes again from the comparison of 
tensions and from studying the spectrum of fluctuations around the 
solution (\ref{sol}).  
\smallskip

The analysis of tensions is identical to the corresponding argument 
in \cite{AlReSc2} and, similar to the case of branes on SU(2), it 
suggests that the brane $(L,M)$ is obtained as a bound state from 
$L+1$ branes of type $(0,M')$ with $M'$ even. On the other hand, the
tension is insensitive to the value $M'$ and hence it does not help
us in deciding which of the branes $(0,M')$ do actually appear. This 
information is encoded partially in the spectrum of fluctuation around our 
solution. Note that the massless states (at $k = \infty$) of the 
configuration (\ref{chain}) come exclusively from open strings 
stretching between two neighbouring branes  $(0,M')$,$(0,M'+2)$ and 
each such pair contributes two massless states. Hence, there are 
$2L$ massless states associated with the chain (\ref{chain}). At 
large but finite $k$, the mass square of these states receives a 
small correction of the form $- 1/k$. If one 
expands the effective action (\ref{redeffact}) around the solution 
(\ref{sol}) one finds that the $2L$ fluctuations which obey the
constraints (\ref{red}) all have the same mass square given by 
$-1/k$. This is in perfect agreement with the spectrum 
of the configuration (\ref{chain}). Even finer information comes 
from the couplings between the branes and the bulk fields 
$\phi_{(l,m)}(z,\bar z)$ using the fact that, to first order 
in $1/k$, these couplings remain unchanged under the RG flow.    
\medskip

Our results can easily be extended to the $N=2$ supersymmetric minimal 
models. The latter are obtained as SU(2)$_k\times$U$_{4}$/U$_{2k+4}$ 
coset theories. Now we need three integers $(l,m,s)$ to label sectors, 
where $l=0,\dots,k$, $m=-k-1,\dots,k+2$ and $s=-1,0,1,2$ are subjected 
to the selection rule $l+m+s=$ even. Maximally symmetric branes 
are labelled by triples $(L,M,S)$ from the same set. We shall restrict 
our attention to the cases with $S=0$. 

The U$_4$ factor in the numerator contributes an additional field X 
which enters the effective action (\ref{redeffact}) minimally coupled 
to the gauge fields $\tA_i, i=1,2$. The solution (\ref{sol}) carries 
over to the new theory if we set X $=0$ and its interpretation is 
the same as for parafermions since the U$_4$ theory remains unperturbed. 
It means once more that a chain of $Q$ 
adjacent ($L =0$)-branes decays into a single ($L=Q-1$)-brane. This 
process admits for  a very suggestive pictorial presentation. Maldacena, 
Moore and Seiberg \cite{MaMoSe} proposed recently to think of 
the target space of parafermion models as a disc with $k+2$ 
equidistant punctures at the boundary labelled by a $k+2$-periodic
integer $q=0,\dots,k+1$. A brane $(L,M)$ is then represented through
a straight line stretching between the points $q_1=M-L-1$ and $q_2
=M+L+1$. In the described process, a chain of branes, each of 
minimal length, decays to a brane forming a straight line between 
the ends of the chain as shown in  Fig.\ 2 (see also \cite{HoIqVa}
for related pictures).
\begin{figure}[h]
\scalebox{.6}{\includegraphics{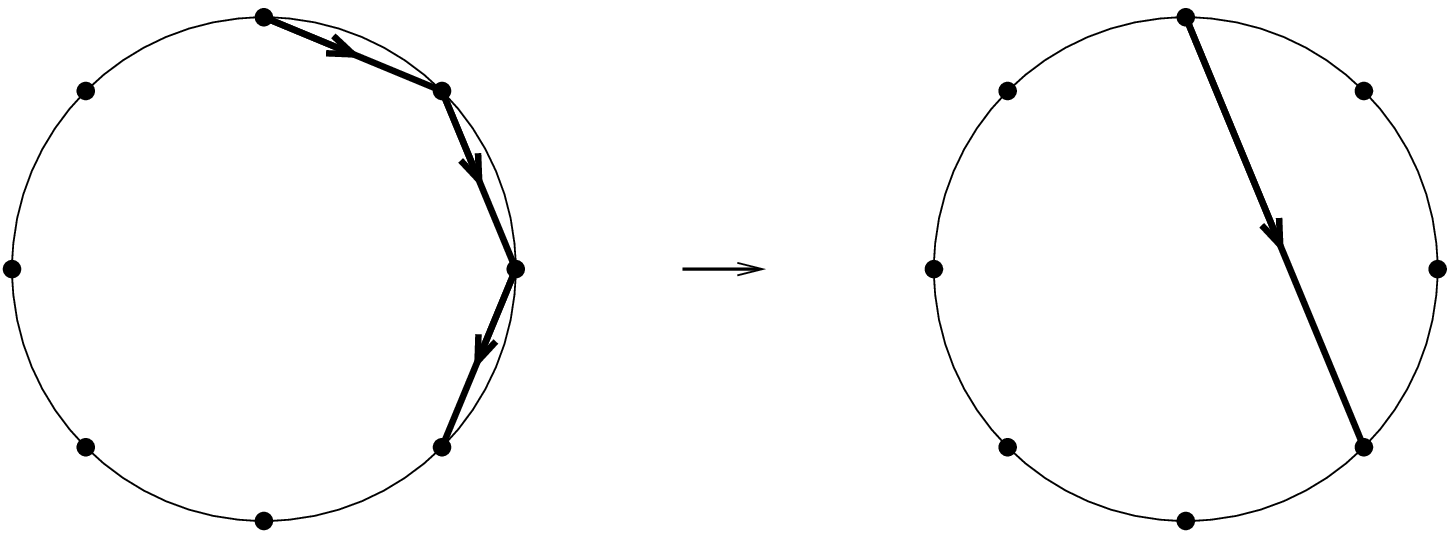}}\vspace*{-2.8cm}
\end{figure} 

\hspace*{9.95cm}\begin{minipage}{3.6cm} {Figure 2:
The chain (\ref{chain}) of branes can 
decay into a single brane $(L,M)$.} \bigskip \bigskip
\end{minipage}

In Figure 2 we have tacitly assumed that the processes we 
identified in the large $k$ regime persist to finite values 
of $k$. For branes on SU(2), analogous results were described in 
Section 4. Similar systematic investigations in case of 
other CFT backgrounds do not exist. But the results
of \cite{WaRu} and the comparison with exact studies 
(see e.g.\ \cite{exact}) in particular models display a 
remarkable stability of the RG flows as we move away from 
the decoupling limit. 
    
\section{Outlook and further directions}

The approach to brane dynamics in non-trivial backgrounds
that we have presented in this note may be extended in 
several different directions. It is certainly possible
to study branes on group manifolds G other than SU(2). Once
more, one can have branes localised along conjugacy 
classes and the above analysis extends to them without 
any difficulties. More interestingly, there exist new 
types of branes whenever the group G admits for non-%
trivial outer automorphism. Their coupling to closed 
string modes was found in \cite{BiFuSc} and it was shown
that they wrap so-called twisted conjugacy classes 
\cite{FFFS}. Constant condensates on stacks of such 
branes where studied in \cite{FreSch} along the lines
of Section 4. It was also  shown there that each of the 
two types of branes contributes its own summand $\Z_x$
to the charge group $C($G$,K)$ and the order $x$ was determined 
for both summands in the case G $=$ SU($N$).  
\smallskip

The groups SU($N$), $N > 3,$ cannot appear directly 
as part of a string background since their central 
charge is too large. But they can show up as  
building blocks of a coset theory with smaller 
central charge. Hence, when combined with suitable
extensions of Section 5, brane dynamics on group 
manifolds other than SU(2) can become relevant for 
string backgrounds. The ideas of Section 5 should be 
directly relevant for branes in Gepner models (see 
\cite{ReSc1}) and for the computations of superpotentials 
that were initiated in \cite{BrSc2}. We plan to return 
to some of these issues in a forthcoming paper.  

\noindent
{\bf Acknowledgements:}  One of us (V.S.) would like to thank A.Yu.\ 
Alekseev and A.\ Recknagel for a very enjoyable collaboration on 
the material presented in Sections 2-4. This research we have 
reported on was supported in parts by the Clay Mathematics 
Institute and by the National Science Foundation under Grant No. 
PHY99-07949. 
 
\def\gaw{Gawedzki}
\newcommand{\sbibitem}[1]{\vspace*{-1.5ex}\bibitem{#1}}

\end{document}